\documentclass[conference]{IEEEtran}
\IEEEoverridecommandlockouts

\usepackage[table]{xcolor}
\usepackage{cite}
\usepackage{amsmath,amssymb}
\usepackage{graphicx}
\usepackage{booktabs}
\usepackage{array}
\usepackage{url}
\usepackage{tikz}
\newcommand{\circlednum}[1]{%
  \tikz[baseline=(X.base)]{
    \node[circle,fill=black,text=white,
          inner sep=1.7pt,minimum size=1.15em,
          font=\scriptsize\bfseries] (X) {#1};
  }%
}
\usepackage{pgfplots}
\pgfplotsset{compat=1.16}
\usetikzlibrary{positioning,shapes.geometric,calc,fit,backgrounds,patterns}
\usepackage[most]{tcolorbox}
\tcbuselibrary{skins,breakable}
\usepackage{pifont}
\usepackage[hidelinks,breaklinks=true]{hyperref}

\newcolumntype{L}[1]{>{\raggedright\arraybackslash}p{#1}}
\newcolumntype{C}[1]{>{\centering\arraybackslash}p{#1}}

\definecolor{revblue}{HTML}{1552C8}

\definecolor{Ink}{HTML}{141414}
\definecolor{SignalBlue}{HTML}{2F69C4}
\definecolor{Paper}{HTML}{FFFFFF}
\definecolor{PaperLight}{HTML}{FAFAFA}

\tcbset{
    summarybox/.style={
        enhanced, breakable=false, sharp corners,
        boxrule=1.1pt, colframe=Ink, colback=Paper,
        title=Summary, coltitle=PaperLight, colbacktitle=Ink,
        fonttitle=\footnotesize\bfseries, halign title=left,
        titlerule=1.8pt, titlerule style={color=SignalBlue},
        lefttitle=3pt, righttitle=3pt, toptitle=1.5pt, bottomtitle=1.5pt,
        left=3pt, right=3pt, top=2pt, bottom=2pt,
        before skip=1.5pt, after skip=1.5pt,
        colupper=Ink, fontupper=\footnotesize
    }
}

\newcommand{\rA}{\rowcolor{gray!12}}
\newcommand{\rB}{\rowcolor{gray!24}}

\definecolor{sBlue}{HTML}{2A78D6}   
\definecolor{sRed}{HTML}{E34948}    
\definecolor{bL250}{HTML}{86B6EF}   
\definecolor{bL350}{HTML}{5598E7}
\definecolor{bL550}{HTML}{1C5CAB}
\definecolor{inkP}{HTML}{0B0B0B}    
\definecolor{inkS}{HTML}{52514E}    
\definecolor{inkM}{HTML}{8A8880}    

\newcommand{\icMag}[1][sBlue]{\tikz[baseline=-0.5ex,scale=0.9]{\draw[#1,line width=0.6pt](0,0)circle(1.05mm);\draw[#1,line width=0.7pt](0.75mm,-0.75mm)--(1.5mm,-1.5mm);}}
\newcommand{\icStack}[1][sBlue]{\tikz[baseline=-0.5ex]{\foreach \y in {0,0.62,1.24}{\draw[#1,line width=0.55pt](-1.1mm,\y mm-0.7mm)--(1.1mm,\y mm-0.7mm);}}}
\newcommand{\icClock}[1][sBlue]{\tikz[baseline=-0.5ex]{\draw[#1,line width=0.55pt](0,0)circle(1.15mm);\draw[#1,line width=0.55pt](0,0)--(0,0.7mm);\draw[#1,line width=0.55pt](0,0)--(0.6mm,0);}}
\newcommand{\icBoard}[1][sBlue]{\tikz[baseline=-0.5ex]{\draw[#1,line width=0.55pt](-1.15mm,-1.15mm)rectangle(1.15mm,1.15mm);\fill[#1](-0.4mm,-0.4mm)rectangle(0.4mm,0.4mm);\draw[#1,line width=0.4pt](-1.15mm,0.6mm)--(-1.7mm,0.6mm);\draw[#1,line width=0.4pt](1.15mm,-0.6mm)--(1.7mm,-0.6mm);}}
\newcommand{\icWave}[1][sBlue]{\tikz[baseline=-0.5ex]{\draw[#1,line width=0.6pt](-1.5mm,-0.7mm)--(-0.7mm,-0.7mm)--(-0.7mm,0.7mm)--(0.1mm,0.7mm)--(0.1mm,-0.7mm)--(0.9mm,-0.7mm)--(0.9mm,0.7mm)--(1.5mm,0.7mm);}}
\newcommand{\icLock}[1][sBlue]{\tikz[baseline=-0.5ex]{\draw[#1,line width=0.5pt](-0.55mm,0.25mm)arc(180:0:0.55mm);\fill[#1](-0.9mm,-1.1mm)rectangle(0.9mm,0.3mm);}}
\newcommand{\icBug}[1][sBlue]{\tikz[baseline=-0.5ex]{\fill[#1](0,0)ellipse(0.8mm and 1.05mm);\foreach \y in {-0.5,0,0.5}{\draw[#1,line width=0.4pt](-0.8mm,\y mm)--(-1.5mm,\y mm+0.25mm);\draw[#1,line width=0.4pt](0.8mm,\y mm)--(1.5mm,\y mm+0.25mm);}}}
\newcommand{\icChip}[1][sBlue]{\tikz[baseline=-0.5ex]{\draw[#1,line width=0.55pt](-1.0mm,-1.0mm)rectangle(1.0mm,1.0mm);\foreach \y in {-0.5,0.5}{\draw[#1,line width=0.4pt](-1.0mm,\y mm)--(-1.55mm,\y mm);\draw[#1,line width=0.4pt](1.0mm,\y mm)--(1.55mm,\y mm);}}}

\newcommand{\icChart}[1][sBlue]{\tikz[baseline=-0.5ex]{\foreach \x/\h in {-0.9/0.7,0/1.3,0.9/2.0}{\fill[#1](\x mm-0.26mm,-1.1mm)rectangle(\x mm+0.26mm,-1.1mm+\h mm);}}}
\newcommand{\icCheck}[1][sBlue]{\tikz[baseline=-0.5ex]{\draw[#1,line width=0.75pt,line cap=round,line join=round](-1.0mm,0.05mm)--(-0.3mm,-0.75mm)--(1.1mm,0.95mm);}}

\newcommand{\icOpen}[1][inkM]{\tikz[baseline=-0.5ex]{\draw[#1,line width=0.6pt](-0.95mm,0)--(0.95mm,0);}}

\begin{document}

\title{Bringing Hardware Security Education to All: A Seamless Curriculum Integration Approach}

\author{\parbox{\textwidth}{\centering\small
Tanvir Hossain\textsuperscript{1}, Mahmudul Hasan\textsuperscript{1}, Kareem El-Araby\textsuperscript{1}, Lorenzo R Godinez\textsuperscript{1}, Swarup Bhunia\textsuperscript{2}, Md Tauhidur Rahman\textsuperscript{3}, Tamzidul Hoque\textsuperscript{1}\\[2pt]
\{tanvir, m.hasan, k035e946, lrgodinez28, hoque\}@ku.edu, swarup@ece.ufl.edu, mdtrahma@fiu.edu\\[2pt]
\textsuperscript{1}EECS, University of Kansas \quad \textsuperscript{2}ECE, University of Florida \quad \textsuperscript{3}ECE, Florida International University}}

\maketitle

\begin{abstract}
Cybersecurity education has predominantly focused on software and network security, under the assumption that the underlying hardware is trustworthy. Threats introduced through the global semiconductor supply chain, together with vulnerabilities arising during design and integration, challenge the assumption that hardware can serve as an unquestioned root of trust. Sensor-enabled systems and machine-learning accelerators further expand the hardware attack surface. The undergraduate curriculum of electrical and computer engineering and computer science programs has not evolved accordingly. Core courses cover the design and the functionality of hardware while the security aspects remain overlooked, and dedicated hardware security (HS) courses are elective, scarce, and often restricted to graduate students. At the University of Florida, fewer than 7\% of the undergraduate electrical and computer engineering students register for the introductory HS course each year. We present an integrated and economical approach that disseminates fundamental HS concepts within existing core courses through contextual learning. Nine hands-on modules were developed such that each security concept exercises an associated design concept of the core course, as a component of that course and not as a course of their own. Doing the security work is how the student exercises the design topic, so the modules are built to strengthen the host course and not only to add to it. Three were assessed across three offerings of the embedded systems course through one course survey and two conducted by an external evaluator, with 25 respondents. We report what those instruments measured and the limitations of the approach.
\end{abstract}


\section{Introduction}
\label{sec:intro}

Electronic devices now hold pivotal information about our surroundings, health, preference, and finance, and they construct decisions on our behalf, which makes their security of the utmost importance. A range of physical and remote invasion techniques using hardware, software, and network interfaces has evolved, and the attack space is proliferating rapidly~\cite{rostami2014primer,trustworthy2025}. The globalization of the semiconductor supply chain has introduced further challenges across design, fabrication, and field operation. Adversaries at an untrusted design or fabrication facility can facilitate piracy, reverse engineering, cloning, and malicious modification of the hardware~\cite{guin2014counterfeit,tehranipoor2017snatchers}. The hardware platform, which serves as the root of trust of a computing system, has thus become the target of many security compromises.

The deployment of machine learning hardware has raised the value of what sits below the software layer. Model architectures have been recovered from shared cache activity on commodity processors~\cite{yan2020cache} and from the electromagnetic emissions of an inference device~\cite{batina2019csi}. A hardware Trojan inserted into an accelerator datapath can steer a classification while the functional test suite remains satisfied~\cite{clements2019hardware}. Language models that emit register transfer level and C code introduce security defects that functional verification does not reject~\cite{pearce2022asleep,thakur2024verigen}. With architecture level attacks such as Meltdown and Spectre~\cite{hill2019spectre}, and software-exploitable hardware bugs that survive conventional verification~\cite{dessouky2019hardfails}, knowledge of hardware security (HS) has become a vital requirement for designing a secure software stack.

The demand for this expertise is measurable. An industry analysis projects that the United States semiconductor workforce will grow from about 345,000 to about 460,000 positions by 2030, and that about 67,000 of the 115,000 new positions are at risk of going unfilled at current degree completion rates, which is 80\% of the projected new technical positions~\cite{sia2023}. Of those unfilled positions, 35\% are engineering or computer science roles at the bachelor's level and a further 26\% require a master's or doctoral degree~\cite{sia2023}. Employment growth for information security analysts is projected at 21\% between 2025 and 2035, against 3\% for all occupations~\cite{bls2025}. The graduates who can work across the two fields come from the same undergraduate programs whose core courses omit the security content.

The undergraduate (UG) curriculum of electrical and computer engineering (ECE), computer science (CS), and related programs has not evolved accordingly. These programs include core courses on the design and the functionality of hardware at various depths, while the security aspects remain predominantly overlooked. The cybersecurity courses offered within the same curricula focus on software and network-oriented attack vectors, and thus fail to create HS expertise. Three obstacles follow from treating HS as a separate course.

\textbf{Limited reach.} Only a few premier institutions have introduced a dedicated course on this topic, and many of those courses are offered to graduate students alone. Where they are available to senior UG students as electives, they are taken by a handful of students who already intend a cybersecurity career with a hardware focus. At the University of Florida, fewer than 7\% of the UG students in the ECE major register for the Introduction to Hardware Security course each year~\cite{nsf2021}. The number of historically underrepresented students taking the course is even lower, which diminishes diversity within the future HS workforce.

\textbf{Cost.} A dedicated HS course requires an HS instructor, laboratory equipment for hands-on learning, classroom resources, and administrative procedures, all of which are difficult for resource-constrained institutions. HS is also a relatively new topic, and few instructors or recent graduates hold the requisite expertise, so a teaching-focused school competes for an instructor it is unlikely to attract. A course taken by a small fraction of the student population is therefore not cost-effective for many institutions.

\textbf{Fragmented view.} A course that presents attacks in isolation separates the attack from the design decision that permitted it. Many attacks originate in the hardware implementation process, side-channel leakage being the familiar example, so a student who meets the attack apart from the design acquires a fragmented view of cybersecurity.

\begin{figure*}[!t]
\centering
\includegraphics[width=\textwidth]{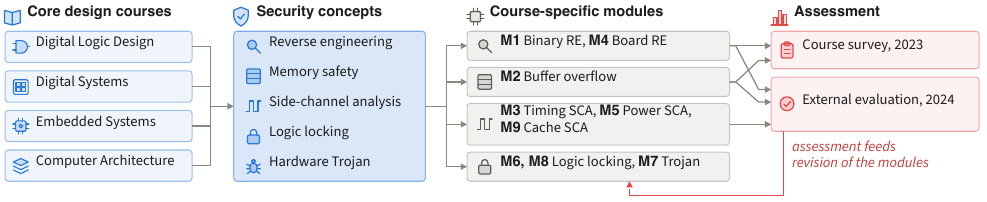}
\caption{The contextual learning framework. Every security concept takes a variant in each core course, attached to the design concept that course already covers. Of the nine modules reported here, M1 to M4 are embedded systems variants, M5 to M8 are digital logic design variants, and M9 belongs to a computer architecture course that the figure does not show. Only M1, M2 and M3 carry assessment data.}
\label{fig:framework}
\end{figure*}

The primary notion behind our solution, outlined in Fig.~\ref{fig:framework}, is to treat HS education as an organic extension of the existing curriculum, and not as a separate, independent offering. Existing theory and laboratory modules within a core course are augmented with security-oriented concepts through a contextual learning process, so that they strengthen the core material while educating on security. The HS concept serves as the tool that exercises the design concept, so that students learn the design and the security aspects of the hardware at the same time. A module is therefore not an extra topic placed beside the syllabus. Performing the attack is how the student works through the design concept, so the security content is built to raise the learning outcome of the host course itself.

The contributions of this paper are as follows.

\begin{itemize}\setlength\itemsep{0.5pt}
\item We identify three obstacles to HS education at the UG level and show that each one follows from treating HS as a separate course (Section~\ref{sec:background}).
\item We present a contextual learning framework that selects each security concept for the core design concept it exercises, and we give five design rules that a module must satisfy (Section~\ref{sec:method}).
\item We develop nine hands-on modules across three core courses that realize five security concepts in our coverage model, and we state which remain unrealized (Sections~\ref{sec:modules} and~\ref{sec:limitations}).
\item We deploy five of the modules and assess three of them across three course offerings, reporting survey data from 25 respondents collected in part by an external evaluator (Section~\ref{sec:results}).
\item We draw four lessons for an adopting institution, covering module status, instructional support, and the defenses a bare-metal target can carry (Section~\ref{sec:discussion}).
\end{itemize}

\section{Background and Related Work}
\label{sec:background}
Hardware security does reach undergraduate students today, through three channels, and it is worth seeing what each can and cannot do before proposing a fourth.

\circlednum{1} The first is the dedicated course and the material written to support it. The community has produced a hands-on textbook built around a board-based laboratory flow~\cite{bhunia2018textbook}, an online laboratory that gives remote access to HS experiments~\cite{hacelab}, and tutorial treatments of topics such as obfuscation and logic locking~\cite{hoque2020tutorial}. \circlednum{2} The second is the competition. Nation-wide contests give teams a design to break under time pressure~\cite{trusthub,csaw}, and workshops at the field's main symposium recruit students into the area~\cite{host}. \circlednum{3} The third channel addresses hardware fundamentals adjacent to security rather than integrating security directly into the curriculum. Examples include configurable platforms and board games~\cite{gameon2024,boardgames2023}, first-year interventions examining hardware-engineering career choice~\cite{career2024}, and inclusive curricula for artificial intelligence of things and intelligent embedded systems~\cite{aiot2024}. These efforts show that hardware topics can reach a broad student population, but they do not incorporate hardware security content.

The first two channels share a property that limits both: each presumes a student who has already chosen the field. The textbook supports a course an institution must first create, the remote laboratory serves an instructor who already teaches the topic, and the competitions draw mostly graduate students. The enrollment figure quoted in Section~\ref{sec:intro} is what that adds up to. These channels therefore primarily reach students who have already developed an interest in hardware security. Their dependence on prior interest limits their ability to broaden participation in the HS workforce.

What is missing is a route that reaches students before they choose. To our knowledge no prior work reports an HS module suite that satisfies three conditions at once: each module is designed around the core design concept it reinforces, the attack is paired with a defensive artifact wherever the target permits one, and the suite is assessed with student survey data from repeated offerings, including an evaluation conducted independently of the instructors. That is the gap this paper addresses.

\section{The Contextual Learning Framework}
\label{sec:method}

The framework rests on a single idea. A security concept belongs in a core design course when it gives the student a reason to use a design concept the course already has to teach. This section describes the benefits of integration, the criterion used to select security concepts, the module-design principles, and the resulting HS coverage.

\subsection{Advantages of Integration}

Integration of HS within existing courses offers four benefits. \circlednum{A1} \textbf{Security education for all}, because core design courses are taken by every student in a program, so the course reaches underrepresented groups automatically and trains core course faculty as HS educators. \circlednum{A2} \textbf{Effective learning}, because many attacks originate in the implementation process, so learning design and security together yields an unfragmented view. \circlednum{A3} \textbf{A sustainable solution}, because the cost of creating and maintaining a new course is removed. \circlednum{A4} \textbf{A future prospect}, because the same contextual learning paradigm can carry other topics with growing workforce demand, such as artificial intelligence and the internet of things~\cite{aiot2024}.

\subsection{The Selection Principle}

A security concept earns a place in a core design course when it exercises a concept the course must teach anyway. Consider the stack frame. An embedded systems course must teach the calling convention, the saved return address, and the position of the stack in the memory map, and students routinely treat all three as bookkeeping. A buffer overflow converts that bookkeeping into a consequence: the student who overruns a buffer and watches the return address change has met the calling convention as the result of a design choice.

The argument runs in the other direction too. A security concept that requires machinery the course does not teach is a new course in disguise. Fault injection on a cryptographic core needs a clock glitcher, a target with a cryptographic accelerator, and a treatment of the cipher, none of which appears in a first embedded systems course, so that topic fails the rule.

\subsection{Design Principles}

Once a security concept satisfies the selection principle, the resulting module follows five design principles. Across the suite, we additionally seek coverage of threats originating both in the supply chain and in the field, while aligning each module with the sequence of its host course.

\textbf{\circlednum{C1} Concept coupling.} The module must exercise a core design concept already in the syllabus. The security topic is chosen for the design concept it reinforces. Its standing in the research literature does not decide the question.

\textbf{\circlednum{C2} Instrument reuse.} The module must use instruments the laboratory already owns and the course already teaches: the debugger, the vendor development environment, and low-cost bench instruments such as the logic analyzer and the digital multimeter. Equipment is the cost that makes a dedicated course impractical, so a module needing new equipment does not transfer to the institutions that need it most.

\textbf{\circlednum{C3} Paired countermeasure.} A module should close with a defense wherever the target permits one. The M2 handout asks the student to state a solution to the overflow, modify the C program accordingly, and explain how the change prevents it. The M3 handout devotes its second part to a constant-time comparison and requires the repaired version to be re-measured with the same five passwords on the same instrument used for the attack. M6 and M8 have the student build the protection itself. M1 as deployed closes with no defense, which we report as a property of that module.

\textbf{\circlednum{C4} Sequence alignment.} The module is placed in the week when the core course covers the concept it exercises. A security module that arrives after the final examination of the related unit has lost its coupling.

\textbf{\circlednum{C5} Split acquisition and analysis.} Some laboratories need more time than the slot allows, so the acquisition step is separated from the analysis step and the analysis is completed outside the laboratory. A board teardown needs the bench, while component identification, trace analysis and waveform interpretation do not.

\begin{figure}[!t]
\centering
\begin{tikzpicture}[font=\scriptsize, x=1cm, y=1cm]
\def\H{0.408}\def\W{4.00}\def\T{5.95}\def\M{8.50}
\node[anchor=west, font=\scriptsize\bfseries, color=inkP] at (0.06,{4*\H+0.17}) {Security concept};
\node[anchor=west, font=\scriptsize\bfseries, color=inkP] at ({\W+0.10},{4*\H+0.17}) {Origin};
\node[anchor=west, font=\scriptsize\bfseries, color=inkP] at ({\T+0.10},{4*\H+0.17}) {Realized as};
\draw[inkM, line width=0.45pt] (0,{4*\H}) -- (\M,{4*\H});
\foreach \r/\ic/\name/\orig/\mod/\done in {%
  3/{\icMag}/{Reverse engineering}/{field}/{M1, M4}/{1},
  2/{\icWave}/{Side-channel analysis}/{field}/{M3, M5, M9}/{1},
  1/{\icLock}/{IP protection, obfuscation}/{supply chain}/{M6, M8}/{1},
  0/{\icBug}/{Hardware Trojan}/{supply chain}/{M7}/{1}}{
  \ifnum\done=1 \fill[sBlue!9] (0,{\r*\H}) rectangle (\M,{\r*\H+\H});\fi
  \draw[inkM!45, line width=0.3pt] (0,{\r*\H}) -- (\M,{\r*\H});
  \node[anchor=west, color=inkP] at (0.06,{\r*\H+0.5*\H}) {\ic{}~~\name};
  \node[anchor=west, color=inkS] at ({\W+0.10},{\r*\H+0.5*\H}) {\orig};
  \ifnum\done=1
    \node[anchor=west, color=inkP] at ({\T+0.10},{\r*\H+0.5*\H}) {\icCheck{}~~\mod};
  \else
    \node[anchor=west, color=inkM] at ({\T+0.10},{\r*\H+0.5*\H}) {\icOpen{}~~\mod};
  \fi}
\draw[inkM, line width=0.45pt] (0,0) -- (\M,0);
\fill[sBlue!9] (0,{-1*\H-0.13}) rectangle (\M,{-0.13});
\draw[inkM!45, line width=0.3pt, dash pattern=on 1.4pt off 1.2pt] (0,{-1*\H-0.13}) rectangle (\M,{-0.13});
\node[anchor=west, color=inkP] at (0.06,{-0.5*\H-0.13}) {\icStack{}~~Memory safety \textit{(addition)}};
\node[anchor=west, color=inkS] at ({\W+0.10},{-0.5*\H-0.13}) {field};
\node[anchor=west, color=inkP] at ({\T+0.10},{-0.5*\H-0.13}) {\icCheck{}~~M2};
\end{tikzpicture}
\caption{The nine modules placed against the security concepts they realize, classified by the stage at which the threat originates. Two of the five concepts originate in the supply chain and two in the field. Memory safety is the fifth and sits outside that split. Section~\ref{sec:limitations} names the concepts the suite does not yet reach.}
\label{fig:coverage}
\end{figure}

\subsection{Coverage Model}
The suite covers five security concepts, four of them classified by the stage at which the threat originates, and Fig.~\ref{fig:coverage} places the nine modules against them. Reverse engineering and side-channel analysis are field threats and were built first. Design obfuscation~\cite{chakraborty2009harpoon,rajendran2015fault} and hardware Trojan insertion~\cite{bhunia2014trojan,xiao2016lessons} originate in the supply chain and are covered by M6, M7, and M8, which also introduce the register transfer level design flow the microcontroller modules do not reach. Memory safety, which the deployed M2 covers, is the fifth, and sits outside the supply chain and field split that orders the other four. Section~\ref{sec:limitations} names the concepts the suite does not yet reach, and we claim no coverage beyond the figure. Breadth is not an end in itself. A course has room for a few modules a semester, so a catalog spanning several concepts and three courses gives an adopting instructor options, and the right choice depends on the course structure, the students' background, and what the instructor can teach.

\subsection{Integration Cost}

Each module adds approximately 30–40 minutes of lecture, for at most about two additional lecture hours across a semester. The remaining adjustment reuses laboratory and assignment slots that already exist. We also cap delivery at three modules a semester, so the security content does not add noticeably to the load the course already places on a student. Where a design course already asks a student to trace an assembly program in a simulator, the reverse engineering module asks the student to trace a program for which no source is supplied. The two tasks occupy the same slot.

\section{Module Design}
\label{sec:modules}

For adoption, each module should follow coverage of the core design topics it exercises. Table~\ref{tab:catalog} gives the nine modules, the design concept each one exercises, the platform, and the graded evidence. Each description introduces the security topic and its significance, recommends course placement based on the supplied materials, and then presents the procedure and student outcome. These placement recommendations are distinct from the deployment history reported below Modules M1, M2, and M3 were delivered in the embedded systems course and carry the survey data of Section~\ref{sec:results}. Module M9 was delivered in the computer architecture course in Fall 2024 as two lecture parts with a recorded measurement, and no survey was administered for it. An early form of M5 was delivered as a graded homework in the embedded systems course in Spring 2022, also without a survey. Modules M4, M6, M7, and M8 exist as written laboratory material only. M6, M7, and M8 were adapted from the laboratory material of the hardware security course at the University of Florida and from a published tutorial on obfuscation and logic locking~\cite{hoque2020tutorial}. Every module, with its laboratory material and lecture recordings, will be released publicly so that other institutions and the community can adopt or adapt it.\footnote{The repository URL is withheld for anonymous review and will be disclosed in the camera-ready version.}

\begin{table*}[!t]
\centering
\caption{Module catalog. Each security concept is selected for the core design concept it exercises. Course: ES is embedded systems, DL is digital logic design, CA is computer architecture. Course labels indicate intended placement; M5 was delivered in embedded systems. Status: D denotes a module delivered with student survey data, T a module delivered without survey data, and W a module that exists as written material only.}
\label{tab:catalog}
\scriptsize
\setlength{\tabcolsep}{3.2pt}
\renewcommand{\arraystretch}{1.02}
\begin{tabular}{|c|L{2.05cm}|c|c|L{1.2cm}|L{4.35cm}|L{2.55cm}|L{3.35cm}|}
\hline
\rowcolor{black}
\textbf{\textcolor{white}{\#}} & \textbf{\textcolor{white}{Module}} & \textbf{\textcolor{white}{Course}} & \textbf{\textcolor{white}{St.}} & \textbf{\textcolor{white}{Origin}} & \textbf{\textcolor{white}{Core design concept exercised}} & \textbf{\textcolor{white}{Platform}} & \textbf{\textcolor{white}{Graded evidence}} \\
\hline
\rA \textbf{M1} & Binary reverse engineering & ES & D & In field & Compilation flow, instruction set, addressing modes, code and data spaces, stack pointer & 8-bit MCU, cross-compiler IDE & Ten answers from the disassembly and memory windows \\
\hline
\rB \textbf{M2} & Buffer overflow & ES & D & In field & Stack frame, calling convention, saved return address, constant initialization & 8-bit MCU, IDE debugger & Input length, secrets leaked, and a justified repair \\
\hline
\rA \textbf{M3} & Timing side channel & ES & D & In field & Timer peripheral, cycle counting, serial port configuration, data-dependent branching & 8-bit MCU, built-in logic analyzer & Timing tables for the vulnerable and repaired versions, compared \\
\hline
\rB \textbf{M4} & Board reverse engineering & ES & W & In field & Microcontroller interfacing, serial protocols, power distribution, datasheets & Populated board, digital multimeter & Bill of materials and block diagram; optionally a schematic and a protocol \\
\hline
\rA \textbf{M5} & Power side channel & DL & T & In field & CMOS switching, output transition and charge transfer, truth tables & Browser-based circuit simulator & Voltage plot, transition table, and candidate truth tables \\
\hline
\rB \textbf{M6} & Combinational logic locking & DL & W & Supply chain & Ripple carry adder, carry propagation, hierarchical design, instantiation & VHDL, FPGA design suite & Locked design, testbench, and waveforms under correct and incorrect keys \\
\hline
\rA \textbf{M7} & Hardware Trojan insertion & DL & W & Supply chain & Decoding a rare input condition, conditional signal assignment, behavioral simulation & VHDL, FPGA design suite & Trigger and payload code, and waveforms dormant and active \\
\hline
\rB \textbf{M8} & Sequential logic locking & DL & W & Supply chain & State machine design, reset and transition behavior, UART frame and transmission & VHDL, FPGA design suite & Augmented state machine, and waveforms for correct and each wrong key \\
\hline
\rA \textbf{M9} & Cache side channel & CA & T & In field & Cache hierarchy, associativity and line size, set indexing, cycle-accurate timing & Commodity x86 host, cycle counter & None; recorded lecture measurement only \\
\hline
\end{tabular}
\end{table*}

\subsection{M1: Binary Reverse Engineering}

\textbf{Overview.} Binary reverse engineering reconstructs program behavior from executable code without access to its source. It enables the inspection of undocumented firmware and reveals how instructions and memory accesses determine a system's behavior.

\textbf{Course integration.} This module fits the program execution unit of an embedded systems course, following coverage of the compilation flow, the 8051 instruction set, addressing modes, memory organization, and function calls. The assignment supplies an 8051 binary and guided questions, supported by lecture slides and recordings on analysis in Keil C51.


\textbf{Procedure.} Students analyze temperature-monitoring firmware that reads stored samples, updates an average, compares it against a threshold, and controls an output port. The exercise guides students through a structured reconstruction of the program execution. Students first identify the program entry point and the return location of the initial function call. They then trace the temperature-reading routine, inspect accesses to external memory, and determine the number, locations, and values of the stored temperature samples. Next, they analyze the arithmetic operations used to compute the average, identify the comparison threshold, and determine which microcontroller output port is affected. These tasks require students to interpret instruction flow, register contents, memory organization, function behavior, and data movement directly from the disassembled binary.

The final part of the module focuses on control-flow analysis. Students examine a branch that returns execution to the beginning of the main routine and consider the effect of replacing it with a function-call operation. This modification causes return addresses to accumulate without corresponding returns, eventually disrupting stack or data memory and altering normal program execution. This task reinforces the relationship between binary-level control flow, function calls, stack behavior, and program correctness.


\begin{tcolorbox}[summarybox, title=Outcome]
The student recovers the sampling rate, the stored values and their locations, the threshold, and the output port from an undocumented binary, and explains the consequence of replacing a conditional jump with a function call. The module closes with no defense, which we report as a property of it.
\end{tcolorbox}

\subsection{M2: Buffer Overflow}

\textbf{Overview.} A buffer overflow occurs when a write exceeds a buffer's bounds and corrupts adjacent memory. Such corruption can redirect execution and expose sensitive data, making memory bounds a security concern in embedded software.

\textbf{Course integration.} The module extends memory organization in embedded systems and should follow C arrays and strings, the 8051 memory map, and function-call and return behavior. Students receive the vulnerable C program and assignment handout, with lecture slides and recordings supporting inspection of memory, control flow, and return values in Keil C51.

\textbf{Procedure.} 
We developed the module around an Intel 8051 program containing an unsafe input-copy operation into a fixed-size memory buffer. The program is intended to return a secret identifier only when the supplied personal information matches a stored record. Students first examine the buffer organization and determine how an input exceeding the allocated space affects neighboring memory. They then construct overlength inputs and observe the resulting execution through the tool's debugger, including changes in memory, control flow, and processor registers. By carefully adjusting the input length and contents, students exploit the overflow to redirect execution to an existing data-return sequence and recover secret values that should not be accessible to the supplied input.

The exercise progressively requires students to recover multiple protected values and to explain how the overflow enables the information leakage. Students then modify the vulnerable program to prevent data from being written beyond the allocated buffer and explain why their modification eliminates the overflow condition. This connects the observed attack behavior with secure memory handling and demonstrates how a seemingly simple input-handling error can lead to unauthorized data disclosure in an embedded system.  The module also introduces common software defenses against memory-corruption attacks, including stack canaries~\cite{cowan1998stackguard} and address space layout randomization (ASLR)~\cite{shacham2004aslr}, allowing students to relate the demonstrated vulnerability to broader mitigation techniques used in modern systems.

\begin{tcolorbox}[summarybox, title=Outcome]
The student leaks successive secrets through the return value register, then writes a repair and justifies it. The two mitigations the lecture names belong to a desktop platform, while the repair the student writes has to hold on a bare-metal 8-bit target.
\end{tcolorbox}

\subsection{M3: Timing Side Channel}

\textbf{Overview.} Timing side channels reveal information through variations in execution time. A password check can therefore return the correct result while still leaking information about the secret through the time taken to reject an input.

\textbf{Course integration.} The module connects timers and serial communication in embedded systems to data-dependent branching. It should follow UART configuration, timer control signals, and interpretation of execution time in machine cycles. The supplied C program, five test inputs, slides, and recordings guide students in measuring the TR0 signal interval with Keil's logic analyzer.

\begin{figure}[!b]\centering
\begin{tikzpicture}
\begin{axis}[
  width=\columnwidth, height=3.28cm,
  ybar,
  bar width=8pt,
  ymin=0, ymax=320,
  ylabel={Machine cycles},
  ylabel style={font=\footnotesize, color=inkS},
  axis line style={inkM, line width=0.4pt},
  tick style={inkM},
  xlabel={Password guess},
  xlabel style={font=\footnotesize},
  symbolic x coords={Howareyou,Baaaa,Aaaaa,Hello,Iello,Iaaaa,Iykpp,Iykyk},
  xtick=data,
  x tick label style={font=\scriptsize, rotate=25, anchor=north east},
  tick label style={font=\footnotesize},
  ymajorgrids, grid style={inkM!35, line width=0.3pt},
  nodes near coords, every node near coord/.append style={font=\tiny, color=inkS},
  enlarge x limits=0.07,
]
\addplot[fill=sBlue, draw=none] coordinates {
 (Howareyou,122) (Baaaa,143) (Aaaaa,152) (Hello,152) (Iello,181) (Iaaaa,188) (Iykpp,240) (Iykyk,291)
};
\end{axis}
\end{tikzpicture}
\caption{\icClock{}~Execution time of the vulnerable password check reported in the M3 lecture materials, in machine cycles, for eight demonstration guesses. The interval is read from the timer run control bit on the logic analyzer. The guesses used in the graded assignment differ from those shown here.}
\label{fig:timing}
\end{figure}
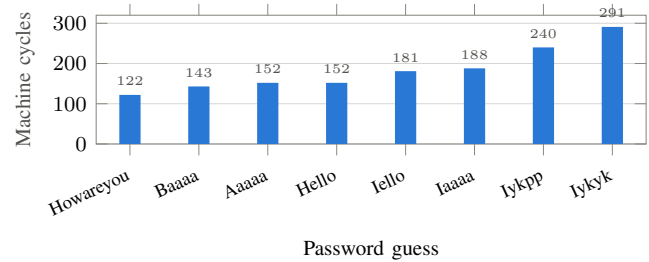

\textbf{Procedure.} The lecture introduces five exploitable side-channel emissions, explains the data dependence of CMOS dynamic power, and connects timing attacks on cryptographic exponentiation to an early-exit string comparison~\cite{kocher1996timing,kocher1999dpa,agrawal2002em}. The module uses an Intel 8051 password-authentication program that compares user input with a stored password and terminates at the first mismatch, creating input-dependent execution times. Fig.~\ref{fig:timing} illustrates this leakage with eight password guesses requiring 122--291 machine cycles.

In the laboratory, students test five password guesses and measure verification time using the debugger's logic analyzer. They then modify the comparison so that all characters are processed regardless of the first mismatch and repeat the measurements. Comparing the vulnerable and constant-time implementations allows students to evaluate the effectiveness of the mitigation using the same instrumentation used to observe the attack.




\begin{tcolorbox}[summarybox, title=Outcome]
Students produce timing tables for both the vulnerable and constant-time implementations and compare the results to evaluate the effectiveness of the mitigation. This is the only module in which the effectiveness of the implemented defense is experimentally measured using the same instrumentation employed to demonstrate the attack.
\end{tcolorbox}

\subsection{M4: Board Reverse Engineering}

\textbf{Overview.} Board reverse engineering identifies the components and infers the functions of an undocumented circuit. Reconstructing this information supports security analysis, debugging, and maintenance when a system's schematic is unavailable.

\textbf{Course integration.} The proposed laboratory fits embedded systems after component identification, datasheet interpretation, and microcontroller interfacing. Its specification calls for an undocumented PCB, a multimeter, and component datasheets. Power distribution and serial protocols provide the background for the optional power-measurement and protocol-identification extensions.

\textbf{Procedure.} The student receives a populated board and no schematic. The required work has four parts. Students identify the board components, construct a bill of materials, infer the function of each board section, and produce a high-level block diagram. Optional work includes continuity mapping from a selected component to the microcontroller, schematic reconstruction, power-line and regulator measurements, and serial-protocol identification. When a logic analyzer is available, students use it to confirm the identified protocol.

\begin{tcolorbox}[summarybox, title=Outcome]
The student produces a bill of materials, a block diagram, and a reasoned functional hypothesis for an undocumented board, which is the first artifact a cloner produces~\cite{kleber2017pcb}. The coverage model classifies reverse engineering as a field threat. The laboratory material specifies no defensive deliverable.
\end{tcolorbox}

\subsection{M5: Power Side Channel}

\textbf{Overview.} Power side-channel analysis infers circuit activity from variations in supply current. Because CMOS switching draws current, internal logic activity can remain observable even when the circuit's output is inaccessible.

\textbf{Course integration.} Delivered in embedded systems in Spring 2022, this exercise also fits digital logic design after truth tables, CMOS switching, and capacitive charging. The assignment provides a Falstad circuit link and input-transition tasks; students infer candidate gate functions from the simulated voltage drop across a supply resistor.

\textbf{Procedure.} The module is delivered in a browser-based circuit simulator. An unknown two-input logic gate sits inside a box, with the two inputs and the supply pin exposed and the output hidden. A sense resistor between the supply pin and the source makes the switching current visible as a voltage. The student toggles the inputs through six ordered transitions, records whether a large voltage transient occurred, builds a transition table, and infers the function of the gate from it.

\begin{tcolorbox}[summarybox, title=Outcome]
The transient reports that an output transition occurred and not its direction, so a gate and its complement yield the same transition table. The student concludes that the circuit is one of two candidates and explains why. The module ends at that analysis and asks for no defense.
\end{tcolorbox}

\subsection{M6: Combinational Logic Locking of a Ripple Carry Adder}

\textbf{Overview.} Combinational logic locking makes a circuit's intended function depend on a key. This laboratory introduces its motivation in intellectual property protection by demonstrating how key-controlled changes to a carry chain affect arithmetic correctness.

\textbf{Course integration.} The laboratory can follow arithmetic circuits in digital logic design, once full adders, ripple carry propagation, VHDL component instantiation, and behavioral simulation have been covered. Students receive a full-adder source file, reference diagrams, and laboratory instructions for Vivado; they construct the adder and develop their own testbench.

\textbf{Procedure.} The student is given only a full adder component in VHDL, builds the 4-bit ripple carry adder hierarchically, then writes a testbench and verifies the arithmetic in behavioral simulation. The carry chain is then locked. The carry signals between stages are replaced by masked carries driven by key-controlled gates, using either XOR and XNOR gates or multiplexers, and a 3-bit key input is added to the entity. For the gate version the reference key is 010, because a key bit of 0 passes a signal through an XOR and a key bit of 1 passes it through an XNOR; for the multiplexer version it is 101. The testbench sweeps all 256 input combinations under a correct key and under incorrect keys.

\begin{tcolorbox}[summarybox, title=Outcome]
Under the correct key the adder matches binary addition across all 256 input combinations. Under an incorrect key the two styles fail differently: the gate version inverts one or more carries, while the multiplexer version substitutes an unrelated signal into the carry chain. In both cases the corruption propagates to the higher bits. The student compares the two styles and reasons about how to enlarge the key space.
\end{tcolorbox}


\begin{figure}[!t]
\centering
\begin{tikzpicture}[font=\scriptsize, x=1cm, y=1cm]
\def\C{1.22}\def\H{0.358}
\foreach \i/\term in {0/{Sp.22}, 1/{Sp.23}, 2/{Sp.24}, 3/{Fa.24}, 4/{later}}{
  \node[font=\scriptsize\bfseries, color=inkP] at ({\i*\C+0.5*\C},{9*\H+0.09}) {\term};}
\foreach \r/\ic/\lab in {0/{\icChip}/{M9 Cache SCA}, 1/{\icLock}/{M8 Seq.\ locking}, 2/{\icBug}/{M7 HW Trojan}, 3/{\icLock}/{M6 Comb.\ locking},
                         4/{\icWave}/{M5 Power SCA}, 5/{\icBoard}/{M4 Board RE}, 6/{\icClock}/{M3 Timing SCA},
                         7/{\icStack}/{M2 Overflow}, 8/{\icMag}/{M1 Binary RE}}{
  \node[anchor=east, font=\scriptsize, color=inkP] at (-0.10,{\r*\H+0.5*\H}) {\ic{}~\lab};}
\foreach \i in {0,1,2,3,4}{\foreach \r in {0,...,8}{
  \draw[inkM!40, line width=0.28pt] ({\i*\C+0.05},{\r*\H+0.025}) rectangle ({\i*\C+\C-0.05},{\r*\H+\H-0.025});}}
\def\cell#1#2#3{%
  \ifnum#3=1 \fill[bL250] ({#1*\C+0.05},{#2*\H+0.025}) rectangle ({#1*\C+\C-0.05},{#2*\H+\H-0.025});\fi
  \ifnum#3=2 \fill[bL350] ({#1*\C+0.05},{#2*\H+0.025}) rectangle ({#1*\C+\C-0.05},{#2*\H+\H-0.025});\fi
  \ifnum#3=3 \fill[bL550] ({#1*\C+0.05},{#2*\H+0.025}) rectangle ({#1*\C+\C-0.05},{#2*\H+\H-0.025});\fi
  \ifnum#3=4 \draw[sBlue!70, line width=0.5pt, dash pattern=on 1.5pt off 1.3pt] ({#1*\C+0.05},{#2*\H+0.025}) rectangle ({#1*\C+\C-0.05},{#2*\H+\H-0.025});\fi
  \ifnum#3=5 \pattern[pattern=north east lines, pattern color=sBlue!60] ({#1*\C+0.05},{#2*\H+0.025}) rectangle ({#1*\C+\C-0.05},{#2*\H+\H-0.025});
             \draw[sBlue!55, line width=0.35pt] ({#1*\C+0.05},{#2*\H+0.025}) rectangle ({#1*\C+\C-0.05},{#2*\H+\H-0.025});\fi}
\cell{0}{4}{1}\cell{0}{6}{1}
\cell{1}{8}{3}\cell{1}{7}{2}
\cell{2}{8}{3}\cell{2}{7}{3}\cell{2}{6}{3}
\cell{3}{8}{5}\cell{3}{7}{5}\cell{3}{6}{5}\cell{3}{0}{5}
\cell{4}{5}{4}\cell{4}{4}{4}\cell{4}{3}{4}\cell{4}{2}{4}\cell{4}{1}{4}
\begin{scope}[yshift=-0.46cm]
\fill[bL250] (0.00,0) rectangle (0.24,0.20);
\node[anchor=west, font=\scriptsize, color=inkS] at (0.28,0.10) {exploratory};
\fill[bL350] (1.54,0) rectangle (1.78,0.20);
\node[anchor=west, font=\scriptsize, color=inkS] at (1.82,0.10) {bonus};
\fill[bL550] (2.72,0) rectangle (2.96,0.20);
\node[anchor=west, font=\scriptsize, color=inkS] at (3.00,0.10) {required};
\pattern[pattern=north east lines, pattern color=sBlue!60] (0.00,-0.30) rectangle (0.24,-0.10);
\draw[sBlue!55, line width=0.35pt] (0.00,-0.30) rectangle (0.24,-0.10);
\node[anchor=west, font=\scriptsize, color=inkS] at (0.28,-0.20) {covered, assignment not stated};
\draw[sBlue!70, line width=0.5pt, dash pattern=on 1.5pt off 1.3pt] (3.56,-0.30) rectangle (3.80,-0.10);
\node[anchor=west, font=\scriptsize, color=inkS] at (3.84,-0.20) {prepared};
\end{scope}
\end{tikzpicture}
\caption{Development history of the modules. The 2022 to 2024 spring columns are taken from the course materials held by the project. The Fall 2024 entries for M1 to M3 are taken from the external evaluation report and the entry for M9 from the computer architecture lecture decks, neither of which states how the topic was assigned. The modules marked prepared carry no survey data.}
\label{fig:evolution}
\end{figure}
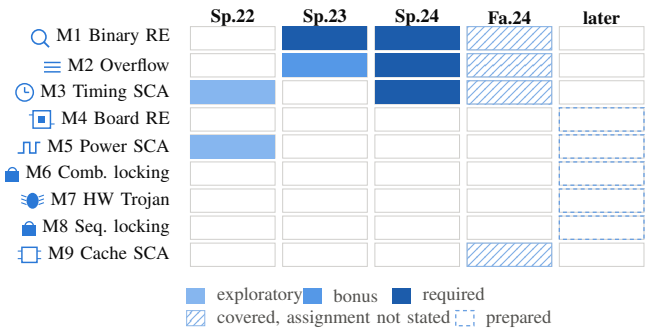

\subsection{M7: Hardware Trojan Insertion}

\textbf{Overview.} This module examines a combinational hardware Trojan that modifies a locking key when a selected operand combination occurs. It illustrates how malicious logic can preserve normal behavior for other inputs while corrupting arithmetic when triggered.

\textbf{Course integration.} A suitable placement is digital logic design after ripple carry adders, VHDL component instantiation, combinational locking, and behavioral simulation, preferably following M6. Trigger and payload construction then exercises decoding and conditional assignment. The supplied full-adder source, locked adder, testbench, and manual confine student changes to the trigger and payload.

\textbf{Procedure.} The student receives the locked adder of M6 with its obfuscation in place and is told not to modify the full adder or the locking logic. The work is confined to a trigger and a payload. The trigger is purely combinational, is built from the two operands alone, and asserts on a single rare input combination. The payload acts only while the trigger is asserted and alters at least one key bit before it reaches the locking gates, the reference solution inverting the lowest. The provided testbench fixes the carry input at 0 and the key at 010 and sweeps all 256 input combinations. The student locates the activation in the waveform and explains how the altered key propagates.

\begin{tcolorbox}[summarybox, title=Outcome]
The design adds correctly for every input except the trigger combination, where the altered key corrupts the carry chain and the sum. The student sees that the locking circuit carries a one-bit payload into a ripple fault across the higher bits, and explains why the Trojan stays hidden for every other input combination.
\end{tcolorbox}

\subsection{M8: Sequential Logic Locking of a UART Transmitter}

\textbf{Overview.} Sequential logic locking requires a circuit to receive the correct key sequence before normal operation begins. It extends design protection to control flow, showing how state transitions can enforce or deny access to a hardware function.

\textbf{Course integration.} The laboratory can extend sequential design in digital logic after VHDL finite state machines, reset behavior, UART framing, and behavioral simulation. Students receive the original UART transmitter source, a starter testbench, reference diagrams, and a manual; they modify the control logic and extend its verification to key sequences.

\textbf{Procedure.} The student examines a working UART transmitter whose finite state machine holds the IDLE, START\_BIT, DATA\_BITS, and STOP\_BIT states, simulates it, and confirms the frame structure. The state machine is then augmented. Three key states and a LOCKED state are added, and the reset state becomes the first key state and no longer IDLE. Three 8-bit key values, AA, BB, and CC in the reference solution, must arrive one per clock while a key valid signal is high, and the comparison runs only then. Any incorrect value sends the machine to LOCKED, which holds the transmit line high with no exit other than reset. The testbench covers the correct sequence and a wrong value at each of the three positions.

\begin{tcolorbox}[summarybox, title=Outcome]
The correct sequence traverses the three key states into IDLE and the transmitter operates normally. Each incorrect value locks the machine at the corresponding stage and the transmit line stays high. The student contrasts a key distributed over several clock cycles with a single combinational key, and explains why the LOCKED state is a blackhole.
\end{tcolorbox}

\subsection{M9: Cache Side Channel}

\textbf{Overview.} Cache side channels infer memory-access activity from the latency difference between cache hits and misses. They show how a mechanism designed to improve performance can also expose information about another process's execution.

\textbf{Course integration.} The attack demonstration fits computer architecture after cache hierarchy, associativity, line size, and set indexing. The first lecture part develops this background, and the second introduces cache attacks and timing examples. The available materials also include a notebook for plotting recorded probe times; no graded assignment is documented.

\textbf{Procedure.} The first lecture part builds the cache from the architecture side, covering the hierarchy from registers to main memory, the attributes that define a cache, and the three associativity types, then has the student read the sizes, associativity and line size of their own machine from the operating system and a topology tool. The second part turns that structure into a channel. It states four requirements for a flush and reload attack, which are shared memory, a high-resolution cycle counter, an instruction that evicts a chosen line, and a last-level cache shared across cores, then walks through flushing a shared line, letting the victim run, and reloading while timing the access. The flush uses the cache line flush intrinsic fenced on both sides, and the reload is timed between two reads of the timestamp counter. A prime and probe variant follows when no memory is shared.

\begin{tcolorbox}[summarybox, title=Outcome]
The lecture closes on a recorded probe-time series separated by a threshold. Across its 90 probes, accesses that hit fall between 96 and 104 cycles and accesses that miss fall between 144 and 168, a gap of 40 cycles around a threshold of 110, which is wide enough to read the victim's access pattern directly off the plot.
\end{tcolorbox}

\section{Deployment and Survey Results}
\label{sec:results}

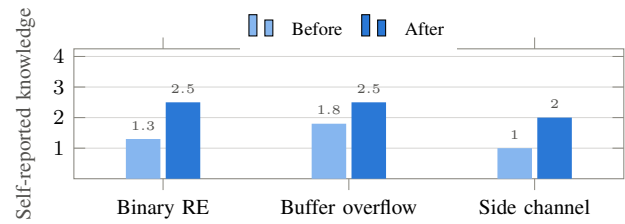
\begin{figure}[!t]
\centering
\begin{tikzpicture}
\begin{axis}[
  width=\columnwidth, height=3.3cm,
  ybar,
  bar width=13pt,
  ymin=0, ymax=4.25,
  ytick={1,2,3,4},
  ylabel={Self-reported knowledge},
  ylabel style={font=\footnotesize, color=inkS},
  axis line style={inkM, line width=0.4pt}, tick style={inkM},
  symbolic x coords={Binary RE, Buffer overflow, Side channel},
  xtick=data,
  tick label style={font=\footnotesize},
  ymajorgrids, grid style={inkM!35, line width=0.3pt},
  legend style={font=\scriptsize, at={(0.5,0.99)}, anchor=south, legend columns=2, draw=none, column sep=5pt, /tikz/every even column/.append style={column sep=5pt}},
  nodes near coords, every node near coord/.append style={font=\tiny, color=inkS},
  enlarge x limits=0.24,
]
\addplot[fill=bL250, draw=none] coordinates {(Binary RE,1.3) (Buffer overflow,1.8) (Side channel,1.0)};
\addplot[fill=sBlue, draw=none] coordinates {(Binary RE,2.5) (Buffer overflow,2.5) (Side channel,2.0)};
\legend{Before, After}
\end{axis}
\end{tikzpicture}
\caption{\icChart{}~Fall 2024 (C3) self-reported knowledge before and after the course on a four-point scale. One is no knowledge and four is strong knowledge.}
\label{fig:knowledge}
\end{figure}

Three of the nine modules have been deployed across sufficient course offerings to support an initial evaluation of student response. What follows gives the deployment record first, then the surveys the numbers come from, and then what those numbers show about ratings, knowledge, outcomes, and the barriers students met.

\subsection{What Was Deployed}

The modules are a component of existing core courses and not a course of their own, and the suite spans three of them. Binary reverse engineering, buffer overflow, the timing side channel, and board reverse engineering attach to an undergraduate embedded systems course; combinational and sequential logic locking, hardware Trojan insertion, and the power side channel to a digital logic design course; the cache side channel to a computer architecture course. Each module enters as lecture segments on units the course already covers, together with the associated homework, and each course keeps its own syllabus, examinations, and grading. The offerings reported here come from the embedded systems and computer architecture courses at the lead institution, whose name and course numbers will be disclosed after review. In the embedded systems offerings the security content accounted for two or three graded assignments, and the weighting changed between offerings. In Spring 2023 the reverse engineering homework carried 10.5 points and the buffer overflow bonus assignment carried 20 points. In Spring 2024 each of the three module assignments carried 2 points.

Fig.~\ref{fig:evolution} gives the development history. The side-channel content began in Spring 2022 as a paper and simulator exercise. Spring 2023 added binary reverse engineering as a required homework and buffer overflow as a bonus assignment. By Spring 2024 all three modules were delivered with complete two-part lecture decks, and buffer overflow had become a required component. The same three topics were taught again in Fall 2024, as the external evaluation report records, and the cache side channel entered the computer architecture course in that term. The suite has grown in every offering since it began. Delivering every module at once was never the aim. At most three ran in any semester, and what each offering taught us about pacing, handout detail, and grading weight was folded into the modules prepared next.

\begin{figure}[!t]
\centering
\begin{tikzpicture}
\begin{axis}[
  width=\columnwidth, height=4.15cm,
  ybar, bar width=5pt,
  ymin=0, ymax=64,
  ylabel={Share (\%)}, ylabel style={font=\footnotesize, color=inkS},
  axis line style={inkM, line width=0.4pt}, tick style={inkM},
  symbolic x coords={Worked ex., Lecture, Materials, Time, TA support},
  xtick=data, x tick label style={font=\scriptsize, rotate=18, anchor=north east},
  tick label style={font=\footnotesize},
  ymajorgrids, grid style={inkM!35, line width=0.3pt},
  legend style={font=\scriptsize, at={(0.5,0.99)}, anchor=south, legend columns=3, draw=none,
                /tikz/every even column/.append style={column sep=4pt}},
  nodes near coords, every node near coord/.append style={font=\tiny, color=inkS, rotate=90, anchor=west},
  enlarge x limits=0.12,
]
\addplot[fill=bL350, draw=none] coordinates {(Worked ex.,33.3) (Lecture,29.2) (Materials,16.7) (Time,12.5) (TA support,8.3)};
\addplot[fill=bL550, draw=none] coordinates {(Worked ex.,28.6) (Lecture,33.3) (Materials,19.0) (Time,14.3) (TA support,4.8)};
\addplot[fill=sRed, draw=none] coordinates {(Worked ex.,0) (Lecture,25) (Materials,25) (Time,25) (TA support,50)};
\legend{{C1, M1},{C1, M2},{C3}}
\end{axis}
\end{tikzpicture}
\caption{\icStack{}~Reported barriers to completion, as a share of selections in Spring 2023 (C1, $n=24$ and $n=21$ selections) and a share of respondents in Fall 2024 (C3, $n=4$). The Spring 2024 cohort was not asked this item. The instrument offered five instructional options and no equipment option.}
\label{fig:barriers}
\end{figure}
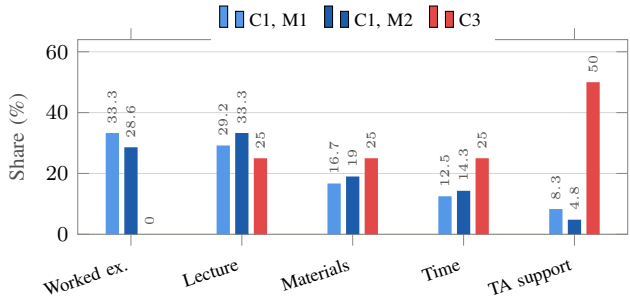
\subsection{Surveys and the External Evaluation}

\begin{table*}[!t]
\centering
\begin{minipage}[t]{0.425\textwidth}
\centering
\caption{Survey cohorts. The dash marks a figure that the source report leaves unfilled. The 2024 reports do not identify the institution of each offering.}
\label{tab:cohorts}
\footnotesize
\setlength{\tabcolsep}{2.6pt}
\renewcommand{\arraystretch}{1.15}
\begin{tabular}{|l|l|l|l|c|c|}
\hline
\rowcolor{black}
\textbf{\textcolor{white}{Cohort}} & \textbf{\textcolor{white}{Term}} & \textbf{\textcolor{white}{Survey}} & \textbf{\textcolor{white}{Modules}} & \textbf{\textcolor{white}{Enr.}} & \textbf{\textcolor{white}{Resp.}} \\
\hline
\rA \textbf{C1} & Spring 2023 & internal & M1, M2 & -- & 13 \\
\hline
\rB \textbf{C2} & Spring 2024 & external & M1, M2, M3 & 13 & 8 \\
\hline
\rA \textbf{C3} & Fall 2024 & external & M1, M2, M3 & -- & 4 \\
\hline
\end{tabular}
\end{minipage}\hfill
\begin{minipage}[t]{0.545\textwidth}
\centering
\caption{Spring 2023 (C1) per-module ratings on a five-point usefulness scale, where the scale midpoint is 3.00.}
\label{tab:c1}
\footnotesize
\setlength{\tabcolsep}{3.5pt}
\renewcommand{\arraystretch}{1.15}
\begin{tabular}{|L{5.35cm}|c|c|c|}
\hline
\rowcolor{black}
\textbf{\textcolor{white}{Item}} & \textbf{\textcolor{white}{Mean}} & \textbf{\textcolor{white}{SD}} & \textbf{\textcolor{white}{N}} \\
\hline
\rA \textbf{M1} usefulness for binary reverse engineering & 2.77 & 1.25 & 13 \\
\hline
\rB \textbf{M1} contribution to assembly and instruction encoding & 2.69 & 1.20 & 13 \\
\hline
\rA \textbf{M1} adequacy of materials and lecture & 2.83 & 1.14 & 12 \\
\hline
\rB \textbf{M2} usefulness for buffer overflow & 2.73 & 1.21 & 11 \\
\hline
\rA \textbf{M2} contribution to memory layout understanding & 2.36 & 1.23 & 11 \\
\hline
\rB \textbf{M2} adequacy of materials and lecture & 2.55 & 1.50 & 11 \\
\hline
\end{tabular}
\end{minipage}
\end{table*}

Three surveys inform this paper. An internal course survey covers the 2023 offering, and an external evaluator administered end-of-course surveys for the two 2024 offerings and issued a combined report covering both, together with a separate summary of the Fall offering.

\textbf{Spring 2023, internal.} A course survey was issued at the end of the term and received 13 responses, with 11 to 13 responses per item. It asked a five-point usefulness scale for each module, a closed item on barriers to completion, and free-response items.

\textbf{Spring and Fall 2024, external.} An external evaluator administered end-of-course surveys for two offerings and issued two reports. The stated purpose was to gather data on the feasibility, the benefits, and the challenges of integrating hardware security concepts within a core design course. The Spring 2024 offering enrolled 13 students and returned 8 responses, a rate of 62\%, and the Fall 2024 report records 4 responses. The evaluator revised the instrument between the two offerings to better reflect the course content, so the four-point knowledge scale and the start value of the interest scale appear only in the Fall report.

\textbf{Respondent profile.} Of the eight Spring 2024 respondents, five were men and three were women, and six identified as Hispanic or Latino. All four Fall 2024 respondents were men, and one identified as Asian. The reports describe their data as coming from two hardware security courses without naming the institution of either, so we report the two 2024 cohorts by term.

\textbf{The external evaluation.} Both 2024 reports were prepared by an evaluator outside the project who holds a doctorate in education and works for an independent evaluation consultancy. The stated purpose was to gather data on the feasibility, the benefits, and the challenges of integrating HS concepts within a core design course. The combined report is organized around four dimensions of broader impact: a stronger workforce, involvement of minority and women students, stimulating UG students in research, and learning through competition. Four findings stand out. Most Spring 2024 students found the modules helped their understanding of embedded systems fundamentals, and both cohorts reported more hands-on learning than in comparable courses. A quarter of the respondents across the two offerings were women, and three quarters of the Spring cohort identified as Hispanic or Latino. Interest in HS rose over the Fall term. On what to strengthen, the barrier data points to teaching assistant support, lecture coverage, and study materials. The items are locally developed Likert and categorical questions, reproduced below without reinterpretation.

\textbf{How the reported values were obtained.} Every value in this section is stated in a source document and is reproduced unchanged, except for four sums and differences of sourced values, which are the combined barrier shares, the respondent total, the assignment point totals, and the knowledge gains. We applied no significance test and pooled nothing across cohorts, because the instruments differ across the three offerings and the samples are too small for a cross-cohort figure to carry meaning. Percentages are given as the source reports them, and counts replace percentages where four or eight responses would imply a precision the sample does not have.

\subsection{Per-Module Ratings}

Table~\ref{tab:c1} gives the Spring 2023 ratings on the five-point usefulness scale. The module means sit between 2.36 and 2.83 on a scale whose midpoint is 3.00, with standard deviations between 1.14 and 1.50. Every item in this cohort therefore sits below the midpoint, and we report that alongside the more favorable 2024 outcomes. The spread is wide in every item, so the responses span the scale. The lowest mean is the contribution of M2 to the prior understanding of memory layout, at 2.36. The free responses explain that result. Several students reported that they did not attempt the assignment, which in that offering carried bonus status, and the module was made required in the following offering.

\subsection{Knowledge and Interest}

Fig.~\ref{fig:knowledge} gives the Fall 2024 knowledge ratings before and after the course on a four-point scale, where one is no knowledge and four is strong knowledge. Binary reverse engineering moved from 1.3 to 2.5, buffer overflow from 1.8 to 2.5, and side-channel attacks from 1.0 to 2.0, with the two largest gains on the topics the cohort knew least at the start. Interest in HS on the ten-point scale moved from 2.8 to 4.5 in the same cohort. Spring 2024 reported interest of 9.1 at the end of the course, alongside an overall experience rating of 7.6 on that scale.

\subsection{Student Outcomes}

Table~\ref{tab:outcomes} gives the categorical outcomes for the two 2024 cohorts. One result holds across both cohorts and one result splits them.

Hands-on exposure is the one result both cohorts agree on. In Spring 2024, 87\% reported more hands-on work through the security modules than in other courses in the program, and in Fall 2024 the figure was 100\%, with no respondent in either cohort disagreeing. That is what an integration should do when it replaces an exercise on the same core concept.

The cohorts part company on whether the modules helped with embedded systems fundamentals, which Spring 2024 reported at 87\% and Fall 2024 at 25\%. We do not smooth that difference away. Two facts in the same instrument explain it: half of the Fall 2024 respondents selected the option stating that they did not attempt the assignments, and half identified teaching assistant support as the improvement they needed most. A module whose assignment is not attempted does not transfer the core concept, which is consistent with the coupling argument of Section~\ref{sec:method}.

Spring 2024 reported stronger downstream outcomes. Participation changed the interest of 75\% of respondents in a graduate degree, every respondent reported looking for an HS internship, and 50\% reported wanting a career in HS. In Fall 2024 no respondent reported a change in graduate degree interest, and the free responses give financial cost and prior plans as the reasons.

\begin{table}[!t]
\centering
\caption{Student outcomes for Spring 2024 (C2) and Fall 2024 (C3). A dash marks an item the revised instrument did not ask.}
\label{tab:outcomes}
\footnotesize
\setlength{\tabcolsep}{4pt}
\renewcommand{\arraystretch}{1.15}
\begin{tabular}{|L{5.25cm}|c|c|}
\hline
\rowcolor{black}
\textbf{\textcolor{white}{Item}} & \textbf{\textcolor{white}{C2}} & \textbf{\textcolor{white}{C3}} \\
\hline
\rA \textbf{More hands-on work than in other courses} & 87\% & 100\% \\
\hline
\rB \textbf{Modules helped embedded systems fundamentals} & 87\% & 25\% \\
\hline
\rA \textbf{Changed interest in a graduate degree} & 75\% & 0\% \\
\hline
\rB \textbf{Sought a hardware security internship} & 100\% & -- \\
\hline
\rA \textbf{Wants a career in hardware security} & 50\% & -- \\
\hline
\rB \textbf{Wants more hardware security modules} & 50\% & -- \\
\hline
\rA \textbf{Needs more time for the assignments} & 75\% & -- \\
\hline
\rB \textbf{Workload higher than similar courses} & 63\% & 25\% \\
\hline
\rA \textbf{Overall experience, ten-point scale} & 7.6 & -- \\
\hline
\end{tabular}
\end{table}

\subsection{Barriers to Completion}

Fig.~\ref{fig:barriers} gives the barriers respondents selected. Every barrier at the top of the list in both cohorts is instructional. In Spring 2023, worked examples and lecture coverage together account for 62.5\% of the selections on M1 and 61.9\% on M2. In Fall 2024 the leading barrier is teaching assistant support at 50\%, with lecture coverage, study materials, and time each at 25\%. The two cohorts disagree on which instructional barrier leads, and they agree on the category. The Spring 2024 cohort was not asked this item.

The instrument offered five options: lecture coverage, study materials, worked examples, time to complete, and teaching assistant support. Laboratory equipment was not among them, so these data cannot rule equipment out. Within the options offered, the response the data supports is more worked examples, longer lecture coverage, and more teaching assistant hours, each cheaper than a dedicated HS course.

\begin{tcolorbox}[summarybox, title=Summary]
Three modules were assessed across three offerings and 25 respondents. In the two cohorts asked, 87\% and 100\% reported more hands-on learning opportunities through the security modules than in other courses, and no respondent disagreed. Self-reported knowledge in Fall 2024 rose by 0.7 to 1.2 points on a four-point scale. The Spring 2023 ratings sit below the scale midpoint, and the reported barriers are instructional in both cohorts asked. The samples are small and self-reported, so these are deployment observations and not statistical findings.
\end{tcolorbox}

\subsection{Sources of the Reported Data}

Every number in this paper comes from a course or evaluation record. The per-module ratings and the 2023 barrier shares come from the Spring 2023 survey export; the interest and knowledge scales, the outcome items, and the 2024 barrier shares from the two evaluation reports; the cycle counts of Fig.~\ref{fig:timing} and the point values from the M3 lecture materials and the assignment handouts; the aims, procedures, outcomes, reference keys, and trigger conditions of M6 to M8 from their laboratory manuals and instructor documents; and M9 with its probe-time figures from the computer architecture lecture decks and the notebook accompanying them. Where a source leaves a field unfilled we mark it as unreported.

\section{Discussion}
\label{sec:discussion}

This section reads the deployment back for what it tells an instructor placing these modules in a course. The free responses, the change in module status, and the reported barriers all point one way: the setting around a module decides how much of it a student receives.

The module that ends in a repair drew the strongest free responses. The 2023 instrument covered M1 and M2 only. One Spring 2023 respondent called the bonus overflow assignment among the most interesting, challenging and rewarding work taken in the program, and another described working out where the return address had to point so that the remaining secrets would leak. M1 carries no defense component, and its responses ask instead for context: one student wrote that it gave a good introduction without saying enough about what the technique is for. The contrast is small, and it runs with the design rule, because the module that ends in a repair ends in something the student built.

That repair also teaches what the desktop lecture cannot. A student who meets the stack canary and address space layout randomization on a server architecture, then works on a target supporting neither, learns that a mitigation is a property of the platform and not of the bug, at no extra cost in a course already working close to the hardware.

Where a module sits in the grade book matters as much as what it contains. The overflow module carried bonus status in Spring 2023 and became required afterwards. One respondent reported being surprised to meet the material on the final examination after skipping an assignment labeled as extra credit. A module chosen because it exercises a core concept is examinable by construction, so leaving it optional sets two signals against each other.

Among the options provided, students most often identified instructional support rather than time as the main barrier. The barrier data in Fig.~\ref{fig:barriers} says so in both cohorts that were asked, and the handouts have already moved that way on their own. Set beside the 2023 buffer overflow handout, the 2024 version adds a partial-credit ladder and procedural hints, among them a note that the return statement shifts with the length of the crafted input. Each of those costs less than a dedicated course.

\section{Limitations}
\label{sec:limitations}

This study developed nine modules across three core courses and gathered student response from three offerings of one of them, and the evidence has clear bounds. The cohorts were not randomized, and no section ran without the modules, so there is no comparison group. The knowledge ratings are self-reported and were not measured with a validated concept inventory, which makes the reported gain a perception of gain. The surveys also sit inside courses with their own examinations, projects and grading, and no instrument separates what the modules contributed from what the rest of the course did.

The samples are small and uneven. The three cohorts returned 13, 8, and 4 responses, a response rate is known only for Spring 2024, at 62\%, and the instrument changed between the 2024 offerings, so several items appear in one cohort only. Counts of this size support no statistical claim. The barrier item was closed and offered no equipment option, so that finding holds only within the options students were given.

Most of the suite is still untested on students. M4, M6, M7, and M8 exist as written laboratory material, and M5 and M9 ran without a survey, so the cache figures rest on a recorded lecture measurement and not on student response. The suite does not yet reach the fabrication stage, physical unclonable functions~\cite{herder2014puf,holcomb2009sram}, bus snooping, fault injection, or gate-level and device-level reverse engineering~\cite{meade2016netlist,quijada2014ic,azriel2021survey}, the last needing decapsulation and imaging equipment that the instrument reuse principle rules out. The three deployed embedded systems modules also share one target, an 8-bit microcontroller reached through a vendor environment that does not run on every student platform, and whether the concepts carry to a 32-bit core is untested.

\section{Conclusion}

Dedicated HS courses reach only a small fraction of undergraduates, and their creation and maintenance are often impractical for resource-constrained institutions. We present a contextual learning approach that treats HS education as an organic extension of the existing curriculum. Nine hands-on modules were developed across embedded systems, digital logic design, and computer architecture, each built so that a security concept exercises a design concept the course must teach anyway, which makes the security content work for the host course as well as the student. The integration costs about two hours of additional lecture across a semester and reuses laboratory and assignment slots that already exist. In the two cohorts given the outcome instrument, 87\% and 100\% of respondents reported more hands-on learning than in other courses, self-reported knowledge rose by 0.7 to 1.2 points on a four-point scale in the cohort with matched ratings, and the external evaluation recorded increased interest in graduate study and HS internships. The result is a delivery route for HS education that does not depend on a student electing the subject. Because core design courses are required, it reaches an entire program, including students that dedicated electives have historically missed, and asks for no new course, instructor line, or equipment. That makes it adoptable where the projected workforce shortfall is least likely to be met, and gives the community a template that travels with the course. We are extending the suite toward the concepts it does not yet reach and testing whether the selection principle carries to a 32-bit target.

\section*{Acknowledgment}
This work was supported in part by the National Science Foundation (NSF) Secure and Trustworthy Cyberspace (SaTC) program under Award No.~2114157. We thank Bradford Davey, Ed.D., of Technology for Learning Consortium, Inc., for conducting the external evaluation.

\makeatletter
\let\IEEEoldthebibliography\thebibliography
\renewcommand{\thebibliography}[1]{%
  \IEEEoldthebibliography{#1}%
  \scriptsize\setlength{\itemsep}{-2.0pt plus 0.2pt}}
\makeatother
\bibliographystyle{IEEEtran}
\bibliography{references}

@inproceedings{yan2020cache,
  author    = {Yan, Mengjia and Fletcher, Christopher W. and Torrellas, Josep},
  title     = {Cache Telepathy: Leveraging Shared Resource Attacks to Learn {DNN} Architectures},
  booktitle = {USENIX Security Symposium},
  pages     = {2003--2020},
  year      = {2020}
}

@inproceedings{batina2019csi,
  author    = {Batina, Lejla and Bhasin, Shivam and Jap, Dirmanto and Picek, Stjepan},
  title     = {{CSI NN}: Reverse Engineering of Neural Network Architectures Through Electromagnetic Side Channel},
  booktitle = {USENIX Security Symposium},
  pages     = {515--532},
  year      = {2019}
}

@inproceedings{clements2019hardware,
  author    = {Clements, Joseph and Lao, Yingjie},
  title     = {Hardware Trojan Attacks on Neural Networks},
  booktitle = {IEEE International Symposium on Circuits and Systems (ISCAS)},
  year      = {2019}
}

@inproceedings{pearce2022asleep,
  author    = {Pearce, Hammond and Ahmad, Baleegh and Tan, Benjamin and Dolan-Gavitt, Brendan and Karri, Ramesh},
  title     = {Asleep at the Keyboard? Assessing the Security of {GitHub} {Copilot}'s Code Contributions},
  booktitle = {IEEE Symposium on Security and Privacy (S\&P)},
  pages     = {754--768},
  year      = {2022}
}

@article{thakur2024verigen,
  author  = {Thakur, Shailja and Ahmad, Baleegh and Pearce, Hammond and Tan, Benjamin and Dolan-Gavitt, Brendan and Karri, Ramesh and Garg, Siddharth},
  title   = {{VeriGen}: A Large Language Model for {Verilog} Code Generation},
  journal = {ACM Transactions on Design Automation of Electronic Systems},
  volume  = {29},
  number  = {3},
  year    = {2024}
}

@article{hill2019spectre,
  author  = {Hill, Mark D. and Masters, Jon and Ranganathan, Parthasarathy and Turner, Paul and Hennessy, John L.},
  title   = {On the {Spectre} and {Meltdown} Processor Security Vulnerabilities},
  journal = {IEEE Micro},
  volume  = {39},
  number  = {2},
  pages   = {9--19},
  year    = {2019}
}

@article{bhunia2014trojan,
  author  = {Bhunia, Swarup and Hsiao, Michael S. and Banga, Mainak and Narasimhan, Seetharam},
  title   = {Hardware {Trojan} Attacks: Threat Analysis and Countermeasures},
  journal = {Proceedings of the IEEE},
  volume  = {102},
  number  = {8},
  pages   = {1229--1247},
  year    = {2014}
}

@article{xiao2016lessons,
  author  = {Xiao, Kan and Forte, Domenic and Jin, Yier and Karri, Ramesh and Bhunia, Swarup and Tehranipoor, Mark},
  title   = {Hardware {Trojans}: Lessons Learned After One Decade of Research},
  journal = {ACM Transactions on Design Automation of Electronic Systems},
  volume  = {22},
  number  = {1},
  year    = {2016}
}

@article{guin2014counterfeit,
  author  = {Guin, Ujjwal and Huang, Ke and DiMase, Daniel and Carulli, John M. and Tehranipoor, Mark and Makris, Yiorgos},
  title   = {Counterfeit Integrated Circuits: A Rising Threat in the Global Semiconductor Supply Chain},
  journal = {Proceedings of the IEEE},
  volume  = {102},
  number  = {8},
  pages   = {1207--1228},
  year    = {2014}
}

@inproceedings{dessouky2019hardfails,
  author    = {Dessouky, Ghada and Gens, David and Haney, Patrick and Persyn, Garrett and Kanuparthi, Arun and Khattri, Hareesh and Fung, Jason M. and Sadeghi, Ahmad-Reza and Rajendran, Jeyavijayan},
  title     = {{HardFails}: Insights into Software-Exploitable Hardware Bugs},
  booktitle = {USENIX Security Symposium},
  pages     = {213--230},
  year      = {2019}
}

@article{rostami2014primer,
  author  = {Rostami, Masoud and Koushanfar, Farinaz and Karri, Ramesh},
  title   = {A Primer on Hardware Security: Models, Methods, and Metrics},
  journal = {Proceedings of the IEEE},
  volume  = {102},
  number  = {8},
  pages   = {1283--1295},
  year    = {2014}
}

@article{tehranipoor2017snatchers,
  author  = {Tehranipoor, Mark M. and Guin, Ujjwal and Bhunia, Swarup},
  title   = {Invasion of the Hardware Snatchers},
  journal = {IEEE Spectrum},
  volume  = {54},
  number  = {5},
  pages   = {36--41},
  year    = {2017}
}

@inproceedings{kocher1996timing,
  author    = {Kocher, Paul C.},
  title     = {Timing Attacks on Implementations of {Diffie-Hellman}, {RSA}, {DSS}, and Other Systems},
  booktitle = {Advances in Cryptology (CRYPTO)},
  pages     = {104--113},
  year      = {1996}
}

@inproceedings{kocher1999dpa,
  author    = {Kocher, Paul and Jaffe, Joshua and Jun, Benjamin},
  title     = {Differential Power Analysis},
  booktitle = {Advances in Cryptology (CRYPTO)},
  pages     = {388--397},
  year      = {1999}
}

@inproceedings{agrawal2002em,
  author    = {Agrawal, Dakshi and Archambeault, Bruce and Rao, Josyula R. and Rohatgi, Pankaj},
  title     = {The {EM} Side-Channel(s)},
  booktitle = {Cryptographic Hardware and Embedded Systems (CHES)},
  pages     = {29--45},
  year      = {2002}
}

@inproceedings{cowan1998stackguard,
  author    = {Cowan, Crispin and Pu, Calton and Maier, Dave and Hinton, Heather and Walpole, Jonathan and Bakke, Peat and Beattie, Steve and Grier, Aaron and Wagle, Perry and Zhang, Qian},
  title     = {{StackGuard}: Automatic Adaptive Detection and Prevention of Buffer-Overflow Attacks},
  booktitle = {USENIX Security Symposium},
  pages     = {63--78},
  year      = {1998}
}

@inproceedings{shacham2004aslr,
  author    = {Shacham, Hovav and Page, Matthew and Pfaff, Ben and Goh, Eu-Jin and Modadugu, Nagendra and Boneh, Dan},
  title     = {On the Effectiveness of Address-Space Randomization},
  booktitle = {ACM Conference on Computer and Communications Security (CCS)},
  pages     = {298--307},
  year      = {2004}
}

@inproceedings{kleber2017pcb,
  author    = {Kleber, Stephan and N{\"o}lscher, Henrik Ferdinand and Kargl, Frank},
  title     = {Automated {PCB} Reverse Engineering},
  booktitle = {USENIX Workshop on Offensive Technologies (WOOT)},
  year      = {2017}
}

@inproceedings{quijada2014ic,
  author    = {Quijada, R. and Raventos, A. and Tarres, F. and Dura, R. and Hidalgo, S.},
  title     = {The Use of Digital Image Processing for {IC} Reverse Engineering},
  booktitle = {IEEE International Multi-Conference on Systems, Signals and Devices (SSD)},
  pages     = {1--4},
  year      = {2014}
}

@inproceedings{meade2016netlist,
  author    = {Meade, Travis and Zhang, Shaojie and Zhao, Zheng and Pan, David and Jin, Yier},
  title     = {Gate-Level Netlist Reverse Engineering Tool Set for Functionality Recovery and Malicious Logic Detection},
  booktitle = {International Symposium for Testing and Failure Analysis (ISTFA)},
  year      = {2016}
}

@article{azriel2021survey,
  author  = {Azriel, Leonid and Speith, Julian and Albartus, Nils and Ginosar, Ran and Shamir, Avi and Paar, Christof},
  title   = {A Survey of Algorithmic Methods in {IC} Reverse Engineering},
  journal = {Journal of Cryptographic Engineering},
  volume  = {11},
  pages   = {299--315},
  year    = {2021}
}

@article{chakraborty2009harpoon,
  author  = {Chakraborty, Rajat Subhra and Bhunia, Swarup},
  title   = {{HARPOON}: An Obfuscation-Based {SoC} Design Methodology for Hardware Protection},
  journal = {IEEE Transactions on Computer-Aided Design of Integrated Circuits and Systems},
  volume  = {28},
  number  = {10},
  pages   = {1493--1502},
  year    = {2009}
}

@article{rajendran2015fault,
  author  = {Rajendran, Jeyavijayan and Zhang, Huan and Zhang, Chi and Rose, Garrett S. and Pino, Youngok and Sinanoglu, Ozgur and Karri, Ramesh},
  title   = {Fault Analysis-Based Logic Encryption},
  journal = {IEEE Transactions on Computers},
  volume  = {64},
  number  = {2},
  pages   = {410--424},
  year    = {2015}
}

@article{hoque2020tutorial,
  author  = {Hoque, Tamzidul and Chakraborty, Rajat Subhra and Bhunia, Swarup},
  title   = {Hardware Obfuscation and Logic Locking: A Tutorial Introduction},
  journal = {IEEE Design and Test},
  volume  = {37},
  number  = {3},
  pages   = {59--77},
  year    = {2020}
}

@article{herder2014puf,
  author  = {Herder, Charles and Yu, Meng-Day and Koushanfar, Farinaz and Devadas, Srinivas},
  title   = {Physical Unclonable Functions and Applications: A Tutorial},
  journal = {Proceedings of the IEEE},
  volume  = {102},
  number  = {8},
  pages   = {1126--1141},
  year    = {2014}
}

@article{holcomb2009sram,
  author  = {Holcomb, Daniel E. and Burleson, Wayne P. and Fu, Kevin},
  title   = {Power-Up {SRAM} State as an Identifying Fingerprint and Source of True Random Numbers},
  journal = {IEEE Transactions on Computers},
  volume  = {58},
  number  = {9},
  pages   = {1198--1210},
  year    = {2009}
}

@book{bhunia2018textbook,
  author    = {Bhunia, Swarup and Tehranipoor, Mark},
  title     = {Hardware Security: A Hands-on Learning Approach},
  publisher = {Morgan Kaufmann},
  year      = {2018}
}

@misc{hacelab,
  title        = {{HACE} Lab: An Online Hardware Security Attack and Countermeasure Evaluation Laboratory},
  howpublished = {\url{https://www.nsf.gov/awardsearch/showAward?AWD_ID=1623310}},
  note         = {{NSF} Award 1623310, accessed Oct.\ 2026},
  year         = {2016}
}

@misc{trusthub,
  title        = {{Trust-Hub}: Hardware Security and Trust Community Resource},
  howpublished = {\url{https://trust-hub.org/}},
  note         = {Accessed Oct.\ 2026}
}

@misc{csaw,
  title        = {{CSAW} Embedded Security Challenge},
  howpublished = {\url{https://www.csaw.io/}},
  note         = {{NYU} Tandon, accessed Oct.\ 2026}
}

@misc{host,
  title        = {{IEEE} International Symposium on Hardware Oriented Security and Trust},
  howpublished = {\url{https://www.hostsymposium.org/}},
  note         = {Accessed Oct.\ 2026}
}

@article{gameon2024,
  author  = {Hossain, Tanvir and Ramirez-Salgado, Adrian and Alsahayan, Eyad and Bhunia, Swarup and Antonenko, Pavlo},
  title   = {Game On: Gamifying Computer Hardware Fundamentals Learning Using a Configurable Platform},
  journal = {IEEE Transactions on Education},
  year    = {2024}
}

@inproceedings{aiot2024,
  author    = {Ramirez-Salgado, Adrian and Bhunia, Swarup and Antonenko, Pavlo and Hwang, Wonjoon and Wusylko, Christine},
  title     = {{NSF IUSE}: Empowering Future Engineers. An Inclusive Curriculum for {AIoT} and Intelligent Embedded Systems},
  booktitle = {ASEE Annual Conference and Exposition},
  year      = {2024}
}

@inproceedings{career2024,
  author    = {Ramirez-Salgado, Adrian and Hossain, Tanvir and Bhunia, Swarup and Antonenko, Pavlo},
  title     = {Supporting Hardware Engineering Career Choice in First-Year Engineering Students},
  booktitle = {ASEE Annual Conference and Exposition},
  year      = {2024}
}

@inproceedings{boardgames2023,
  author    = {Ramirez-Salgado, Adrian and Hossain, Tanvir and Hoque, Tamzidul and Bhunia, Swarup and Koroly, Mary Jo and Davey, Bradford},
  title     = {Engaging Students in Exploring Computer Hardware Fundamentals Using {FPGA} Board Games},
  booktitle = {ASEE Annual Conference and Exposition},
  year      = {2023}
}

@article{trustworthy2025,
  author  = {Prabhakara Rao, Shams and Balaji, Sai Ram Abhinav and Rahman, Fahim and Suha, Tasnuva and Mahfuz, Tanzim and Hossain, Tanvir},
  title   = {Secure and Trustworthy Microelectronics: Vulnerabilities, Solutions, and Trends},
  journal = {Journal of Hardware and Systems Security},
  volume  = {9},
  number  = {3},
  pages   = {163--189},
  year    = {2025}
}

@misc{sia2023,
  author       = {{Semiconductor Industry Association}},
  title        = {Chipping Away: Assessing and Addressing the Labor Market Gap Facing the {U.S.} Semiconductor Industry},
  howpublished = {\url{https://www.semiconductors.org/chipping-away-assessing-and-addressing-the-labor-market-gap-facing-the-u-s-semiconductor-industry/}},
  year         = {2023},
  note         = {Accessed Oct.\ 2026}
}

@misc{bls2025,
  author       = {{U.S. Bureau of Labor Statistics}},
  title        = {Occupational Outlook Handbook: Information Security Analysts},
  howpublished = {\url{https://www.bls.gov/ooh/computer-and-information-technology/information-security-analysts.htm}},
  year         = {2025},
  note         = {Employment projections 2025--2035, accessed Oct.\ 2026}
}

@misc{nsf2021,
  author       = {{Hoque}, Tamzidul},
  title        = {Collaborative Research: {SaTC}: {EDU}: Hardware Security Education for All Through Seamless Extension of Existing Curricula},
  howpublished = {NSF Award Number 2114157, Directorate for STEM Education, Division of Graduate Education, \url{https://ui.adsabs.harvard.edu/abs/2021nsf....2114157H}},
  month        = jul,
  year         = {2021}
}

\end{document}